\documentclass[a4paper,10pt]{article}
\usepackage{amssymb}
\usepackage{graphicx}
\usepackage{cite}

\let\OLDthebibliography\thebibliography
\renewcommand\thebibliography[1]{
  \OLDthebibliography{#1}
  \setlength{\parskip}{2pt}
  \setlength{\itemsep}{0pt plus 0.3ex}
}

\def\nn{\nonumber}

\def\arcsinh{\mathop{\mbox{arcsinh}}}
\def\diag{\mathop{\mbox{diag}}}

\begin{document}

\title{Wormhole solutions of RDM model}
\author{Igor Nikitin\\Department of High Performance Analytics\\Fraunhofer Institute for Algorithms and Scientific Computing\\Schloss Birlinghoven, 53757 Sankt Augustin, Germany\\ \\ igor.nikitin@scai.fraunhofer.de}
\date{}

\maketitle

\begin{abstract}
\noindent The model of a spiral galaxy with radially directed flows of dark matter is extended by exotic matter, in a form of a perfect fluid with a linear anisotropic equation of state. The exotic matter is collected in the minimum of gravitational potential and opens a wormhole in the center of the galaxy. The flows of dark matter pass through the wormhole and form a mirror galaxy on the other side. The influence of model parameters to the shape of solution is studied, a solution matching parameters of Milky Way galaxy is computed.
\end{abstract}

\vfill

\begin{figure}[h]
\centering
\includegraphics[width=0.65\textwidth]{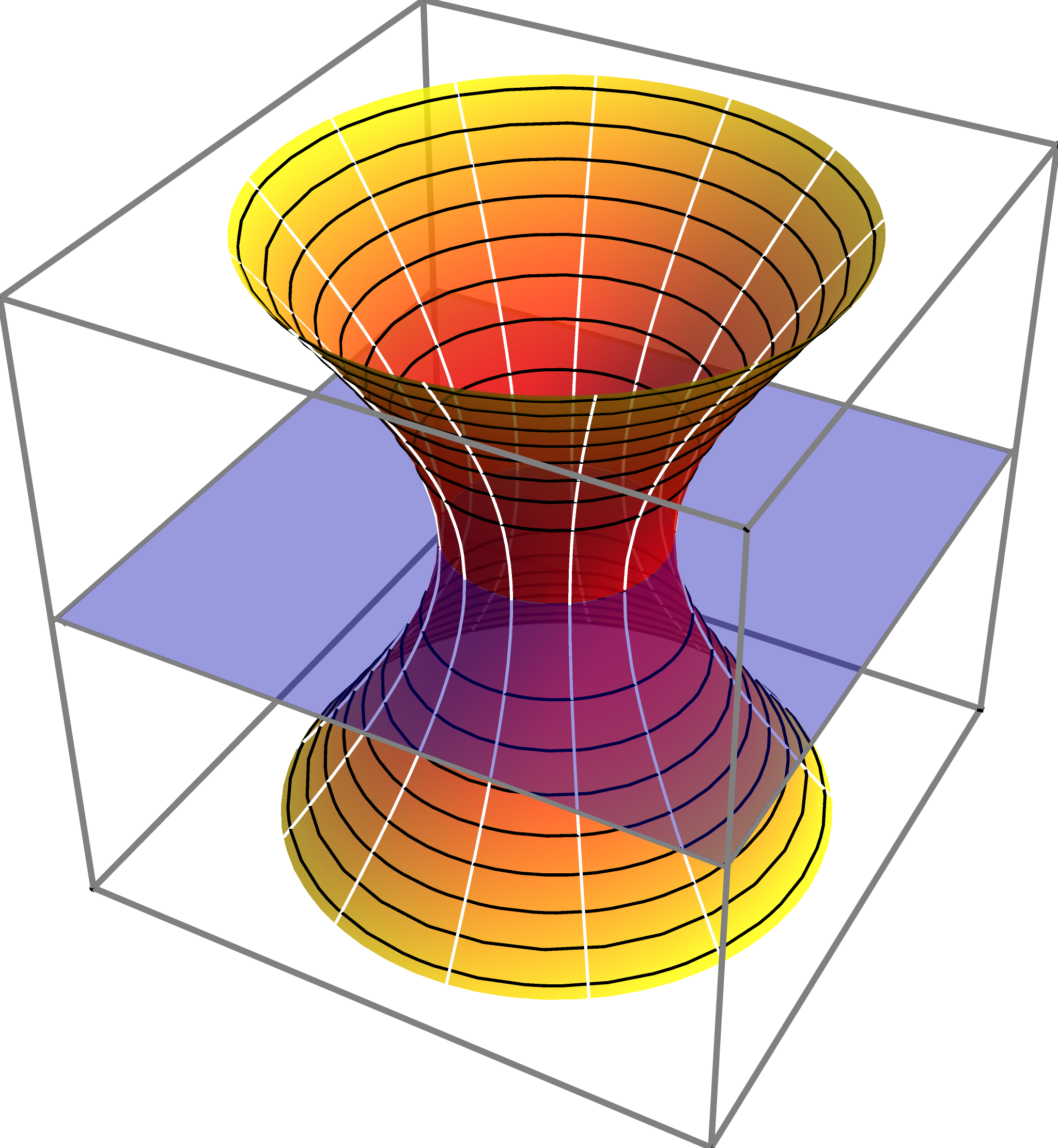}
\caption{Wormhole solution. Horizontal axes are two spatial coordinates $x,y$, vertical axis is a proper length $L$. Color represents a redshift function. White curves show the world lines of dark matter in RDM model.}
\label{f1}
\end{figure}

\newpage

\hfill With kind acknowledgement to my father, Nikolay Nikitin, 

\hfill for fascinating me with warp drives, wormholes and time machines.

\section{Introduction}\label{intro}

In the present paper we continue investigation of a model, formulated in \cite{tachyo-dm,static-rdm}, considering a spiral galaxy with dark matter flows radially converging to the center of the galaxy. Such geometry leads to flat rotation curves at large distances, almost Keplerian orbits of  S-stars, surrounding the central massive dark object, further, in strong fields, solution goes towards Schwarzschild regime. Unexpected differences start near gravitational radius. The flows of dark matter prevent creation of event horizon, instead producing a spherical region with extremely large redshift. 

In this paper we will study the modifications of the model that open a wormhole in the center of galaxy. In this model the world lines of dark matter pass through the wormhole to the other side, in a parallel universe or in a distant part of our one. On the other side the flows of dark matter radially diverge and a mirror galaxy is formed. Fig.\ref{f1} illustrates the related geometry of space-time and direction of dark matter flows. A possibility that the massive dark object in galactic center is a wormhole, has been previously considered in papers \cite{kardashev,wormhole1,wormhole2}.

According to general theory of wormholes \cite{visser}, these solutions require exotic matter, that is the matter with negative sum of density and pressure $\rho+p<0$. The dark matter, considered in our model, is not exotic in this sense. By its composition it can be massive, null or tachyonic, but it has $\rho>0$, $p=0$, a normal dust type. Therefore, for our purpose an explicit extension of the model by exotic matter is necessary.

We have tried various extensions and will present here the successful one, an exotic perfect fluid, possessing a linear anisotropic equation of state. This is certainly not a unique choice. Other possibilities include magnetic field passing through the wormhole \cite{kardashev}, thin shells of exotic matter \cite{visser,thinshell2}, modified Chaplygin gas \cite{wrmh-chapl-gas1,wrmh-chapl-gas2}, other mechanisms \cite{wrmh-th1,wrmh-th2,wrmh-th3,wrmh-th4,wrmh-th5,wrmh-th6,wrmh-th7,wrmh-th8,wrmh-th9}. Any of these mechanisms can be used to open a wormhole, then we just need to thread it with the dark matter flows and find self-consistent solution. In this paper we will concentrate on static spherically symmetric solutions and postpone a detailed study of stability and dynamics to the other occasion.

The first concern comes from the following property of wormholes, mentioned in \cite{visser}. The matter, passing through the wormhole in one direction, leads to dynamical increase of wormhole mass from one side and decrease from the other side. Fortunately, our dark matter flows come in pairs, as energetically balanced combination of incoming and outgoing flows. In this way the total mass of central object is conserved and static wormhole solutions become possible.

In Section~\ref{sec2} we derive main equations, describing geodesic flows of dark matter, hydrostatic equilibrium of exotic fluid and gravitational field equations. In Section~\ref{sec3} we describe numerical procedures used for solution of the equations. In Section~\ref{sec4} we select model parameters, appropriate for the dimensions of Milky Way galaxy.

\section{The equations}\label{sec2}

Consider static spherically symmetric ({\it S3\,}) space-time in spherical coordinates $(t,r,\theta,\phi)$ with a standard line element:
\begin{equation}
ds^2=-A(r)dt^2+B(r)dr^2+r^2(d\theta^2+\sin^2\theta\; d\phi^2).
\end{equation}

Dark matter distribution is composed of two $T$-symmetric outgoing / ingoing radial dark matter flows, described by density $\rho(r)$ and velocity vectors $u_\pm(r)=(\pm u^t(r),u^r(r),0,0)$. Each flow satisfies a system of geodesic equations, for which in \cite{static-rdm} a general solution in arbitrary {\it S3\,}-metric has been found: 
\begin{eqnarray}
&&\rho=c_1/\left(r^2u^r\sqrt{AB}\right),\ u^t=c_2/A, \ u^r=\sqrt{c_2^2+c_3A}/\sqrt{AB}.\label{eq_geode}
\end{eqnarray}
Here $c_{1-3}$ are arbitrary constants. For normal (non-exotic) type of matter $c_1>0$. The sign of $c_2$ exchanges ingoing/outgoing flows and can be fixed to $c_2>0$ by convention. The third constant defines a covariant norm $c_3=u_\mu u^\mu$ and can be set to three discrete values $c_3=-1,0,1$, dependently on the matter type, massive, null or tachyonic.

The first possibility to add exotic matter is to allow for $c_1<0$, i.e., $\rho<0$ in RDM model. However, this will create a globally antigravitational contribution of dark matter and the galaxy will not be formed. 

The next possibility we have considered is an extension of the model with additional dark matter flows. The formula (\ref{eq_geode}) will remain valid if multiple layers are introduced, each including $T$-conjugated pair of dark matter flows, described by own $c_{1-3}$ constants. These layers contribute linearly to energy-momentum tensor and the resulting field equations. The choice of coefficients can be managed to support correct sign of gravity at large distance and exotic type of matter near  the center. Our numerical experiments show that it's indeed easy to select the constants opening a wormhole in the center. However, this model does not allow to control efficiently the behavior of solution on the other side of the wormhole. 

We have performed Monte Carlo numerical experiments with 2-and 3-layered RDM model, randomly selecting parameters in a wide range, generating 1 million of solutions for each configuration of layers. We have found no wormhole solutions with the required asymptotics from both sides. The solutions always end either in naked singularity $A\to\infty$, $B\to0$, or in unidentified phenomenon with $A\to0$, $B\to0$. Although it would be interesting to clarify what exactly this phenomenon is, currently it is not what we are looking for. We remind that event horizon has $A\to0$, $B\to\infty$, while the wormhole corresponds to a finite positive limit of $A$ and $B\to\infty$.

As the next attempt, we select a perfect fluid model for exotic matter. The fluid is described by density $\rho_{exo}$ and two pressures $p_r$, $p_t$ of radial and transverse type. They satisfy hydrostatic equation: 
\begin{eqnarray}
&&r (p_r+ \rho_{exo}) A'_r + 2 A\ (r (p_r)'_r + 2 p_r - 2 p_t)=0.
\end{eqnarray}
The equation can be easily derived by considering covariant conservation condition  $\nabla_\mu T^\mu_\nu=0$ for energy-momentum tensor $T^\mu_\nu=\diag(-\rho_{exo},p_r,p_t,p_t)$.

Further, selecting a linear anisotropic equation of state (EOS):
\begin{eqnarray}
&&\rho_{exo}=k_1 p_r,\ p_t=k_2 p_r,
\end{eqnarray}
described by two constants $k_{1,2}$, find a general solution in arbitrary {\it S3\,}-metric:
\begin{eqnarray}
&&p_r=k_3 A^{k_4} r^{k_5},\ k_4=-(1 + k_1)/2,\ k_5=2 (-1 + k_2). 
\end{eqnarray}
The constants $k_{1,2}$ here come from EOS, the constant $k_3$ influences integral values, measuring the total amount of exotic matter.

Analyzing a large collection of wormhole solutions, given in \cite{visser}, we see that $p_t>0$, $p_r<0$ are generally preferred. Also, to ensure unconditional satisfaction of $\rho_{exo}+p_r<0$, we require that $\rho_{exo}<0$. This leads to {\it a priori} conditions $k_1>0$, $k_2<0$, $k_3<0$ we impose on the model.

Considering the behavior of solution at $r\to\infty$, $A\to1$, we see that $p_r,p_t,\rho_{exo}\to0$ at $k_5<0$, i.e., $k_2<1$. For our choice of coefficients this condition is automatically satisfied.

Considering further the limit $r\to0$, $A\to0$, we have $p_r\to-\infty$ at $k_4<0$, $k_5<0$, i.e., $k_1>-1$, $k_2<1$. Again,  our choice of coefficients satisfies this condition and provides a reasonable behavior of solution: an accumulation of large negative $p_r$ and $\rho_{exo}$ in deep minima of $A$, available in RDM model. Like a normal fluid is collecting in a minimum of gravitational potential, the exotic fluid also does so. Although its mass is negative and gravitational force is opposite to the direction of gravitational field, its acceleration is directed along the field, providing the behavior similar to the normal fluid. Only the own field created by this fluid has characteristic antigravitational behavior. In strong field limit it serves as a mechanism for creating naked singularities and opening wormholes.

We note also that the presence of transverse pressure $p_t$, controlled by $k_2$ parameter, serves as an important feature for further adjustments of the model. The absence of this parameter in the multilayered extension of RDM model is a possible reason for the absence of wormhole solutions with the necessary asymptotics in this model.

Einstein field equations have a form:
\begin{eqnarray}
&&A'_r=(A(1 - B^{-1}) + c_4 \sqrt{1 + c_5 A}- exo ) B/r,\nn\\  
&&B'_r=(-A(1 - B^{-1}) + c_4 /\sqrt{1 + c_5 A}- k_1 exo) B^2/(Ar),\\
&&c_4=4c_1c_2,\ c_5=c_3/c_2^2,\ exo=-2Ar^2 p_r=-2k_3A^{k_4+1}r^{k_5+2}.\nn
\end{eqnarray}
There are also other equations, associated with angular components, which follow from these two and are satisfied automatically. Geodesic equations for RDM flows are equivalent to covariant conservation conditions for energy-momentum tensor, written separately for every flow. Together with similar conditions for exotic matter, they lead to covariant conservation of total energy-momentum tensor. It serves as compatibility condition for the whole system of field equations \cite{dirac,blau}. As a result, two angular components of field equations are automatically satisfied on the surface of two other field equations, geodesic and hydrostatic equations. We have also verified the validity of angular equations by straightforward computation.

One can relate the constants $c_{4,5}$ with the other constant, defining a square of orbital velocities at large distances:
\begin{eqnarray}
&&\epsilon=(\sqrt{1 + c_5}+1/\sqrt{1 + c_5})\ c_4/2.
\end{eqnarray}
This constant is measurable from galactic rotation curves \cite{static-rdm}. For Milky Way galaxy it equals $\epsilon\sim4\cdot10^{-7}$. Instead of specifying $c_{4,5}$, one can set $\epsilon,c_5$ and reconstruct $c_4$ from this formula.

If one starts the integration at large distances, the exotic term can be safely neglected. Here the weak field theory \cite{tachyo-dm} becomes applicable and solutions have a form:
\begin{eqnarray}
&&a=Const-r_s e^{-x} +2\epsilon x,\ b=c_4/\sqrt{1 + c_5} +r_s e^{-x} \label{asympt}
\end{eqnarray}
in logarithmic variables
\begin{eqnarray}
&& x=\log r,\ a=\log A,\ b=\log B. \label{xab}
\end{eqnarray}
The starting value $a_1=0$ can be chosen by convention, meaning that global time is measured by the clock of the observer, located at initial distance $r_1$. This setting fixes the additive constant in the asymptotic formula above. The starting value $b_1$ is bound by this formula to the constant $r_s$, Schwarzschild's gravitational radius.

For studying wormhole solutions, it is convenient to introduce a new variable $B=Z^{-1}$ before proceeding to logarithmic variables. This transforms wormhole throat position to finite domain, from $B\to\infty$ to $Z=0$. The equations become
\begin{eqnarray}
&&A'_r=(A(1 - Z) + c_4 \sqrt{1 + c_5 A}- exo )/(Zr),\\  
&&Z'_r=(-A(1 - Z) + c_4 /\sqrt{1 + c_5 A}- k_1 exo) / (-Ar).\nn
\end{eqnarray}
Further, transforming the variable to $Z=W^2$, one can allow change of sign, $W>0$ before the throat, $W<0$ after the throat:
\begin{eqnarray}
&&A'_r=(A(1 - W^2) + c_4 \sqrt{1 + c_5 A}- exo ) /(W^2r),\\  
&&W'_r=(-A(1 - W^2) + c_4 /\sqrt{1 + c_5 A}- k_1 exo) / (-2AWr).\nn
\end{eqnarray}
Let us introduce an element of proper length $dL=\pm\sqrt{B}dr=W^{-1}dr$. Crossing the throat, we see that $dr<0$ before the throat, $dr>0$ after the throat, while $dL$ preserves the sign. The equations become
\begin{eqnarray}
&&r'_L=W,\nn\\
&&A'_L=(A(1 - W^2) + c_4 \sqrt{1 + c_5 A}- exo ) /(Wr),\\  
&&W'_L=(-A(1 - W^2) + c_4 /\sqrt{1 + c_5 A}- k_1 exo) / (-2Ar).\nn
\end{eqnarray}
The same result can be obtained by direct derivation of field equations with line element
\begin{equation}
ds^2=-A(L)dt^2+dL^2+r^2(L)(d\theta^2+\sin^2\theta\; d\phi^2).
\end{equation}
Now we can proceed to logarithmic variables (\ref{xab}), introducing additionally
\begin{equation}
w=\arcsinh W,\ k_6=\log(-k_3).
\end{equation}
New variable $w$ allows for sign change of $W$ and behaves logarithmically, similar to $b$, in the limit of large $W$, i.e., small $B$. The logarithmic transform of $k_3$ constant simplifies further formulae. The equations become
\begin{eqnarray}
&&x'_L=\sinh w\ /e^{x},\ exo=2e^{ k_6 + a (k_4+1) + x (k_5+2)},\nn\\
&&a'_L=(e^a(1 - \sinh^2 w) + c_4 \sqrt{1 + c_5 e^{a}}- exo )/(\sinh w\ e^{a+x}),\label{daL}\\  
&&w'_L=(-e^a(1 - \sinh^2 w) + c_4 /\sqrt{1 + c_5 e^{a}}- k_1 exo) / (-2 \cosh w\ e^{a+x}).\nn
\end{eqnarray}

\begin{figure}
\centering
\includegraphics[width=0.49\textwidth]{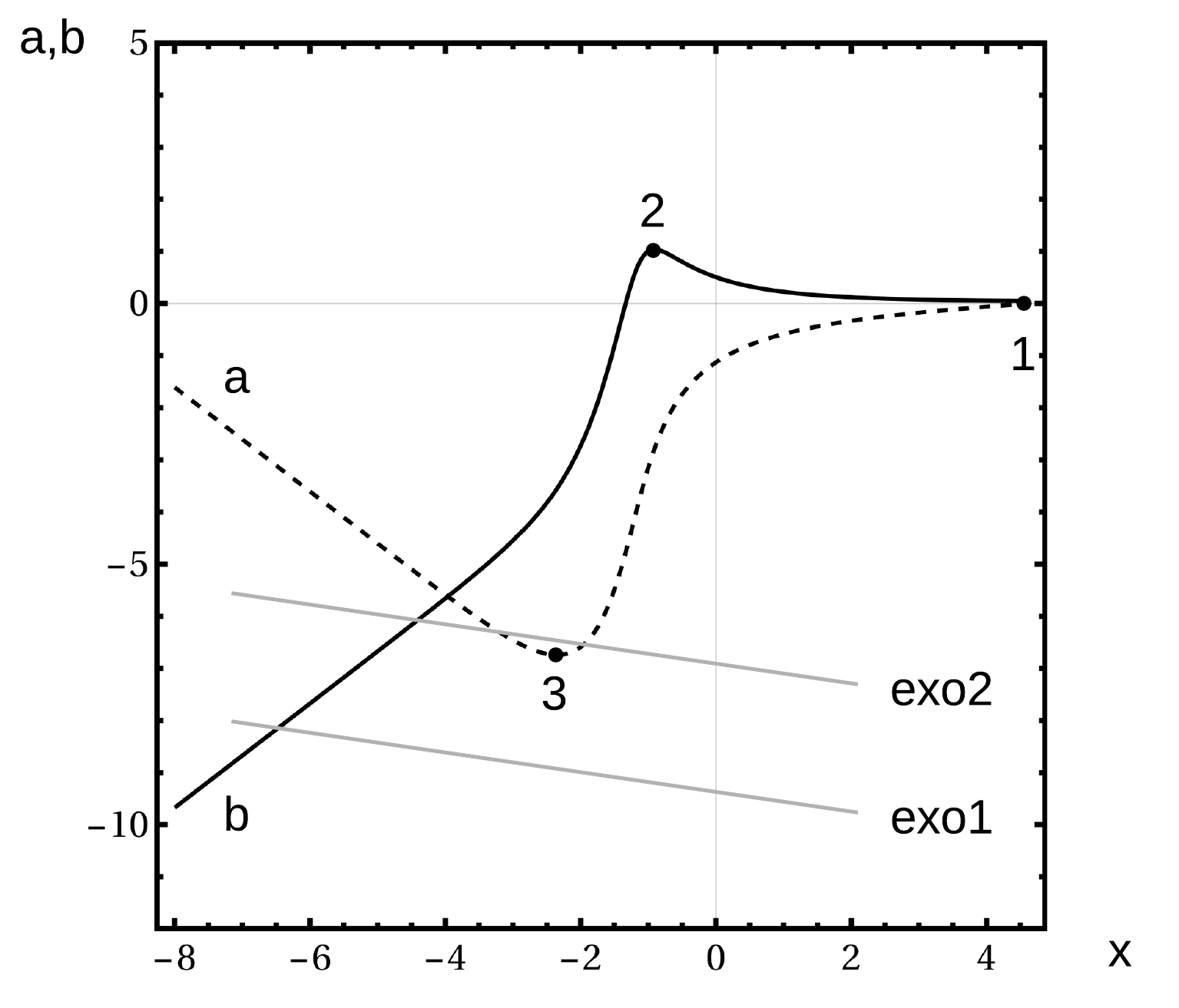}
\includegraphics[width=0.49\textwidth]{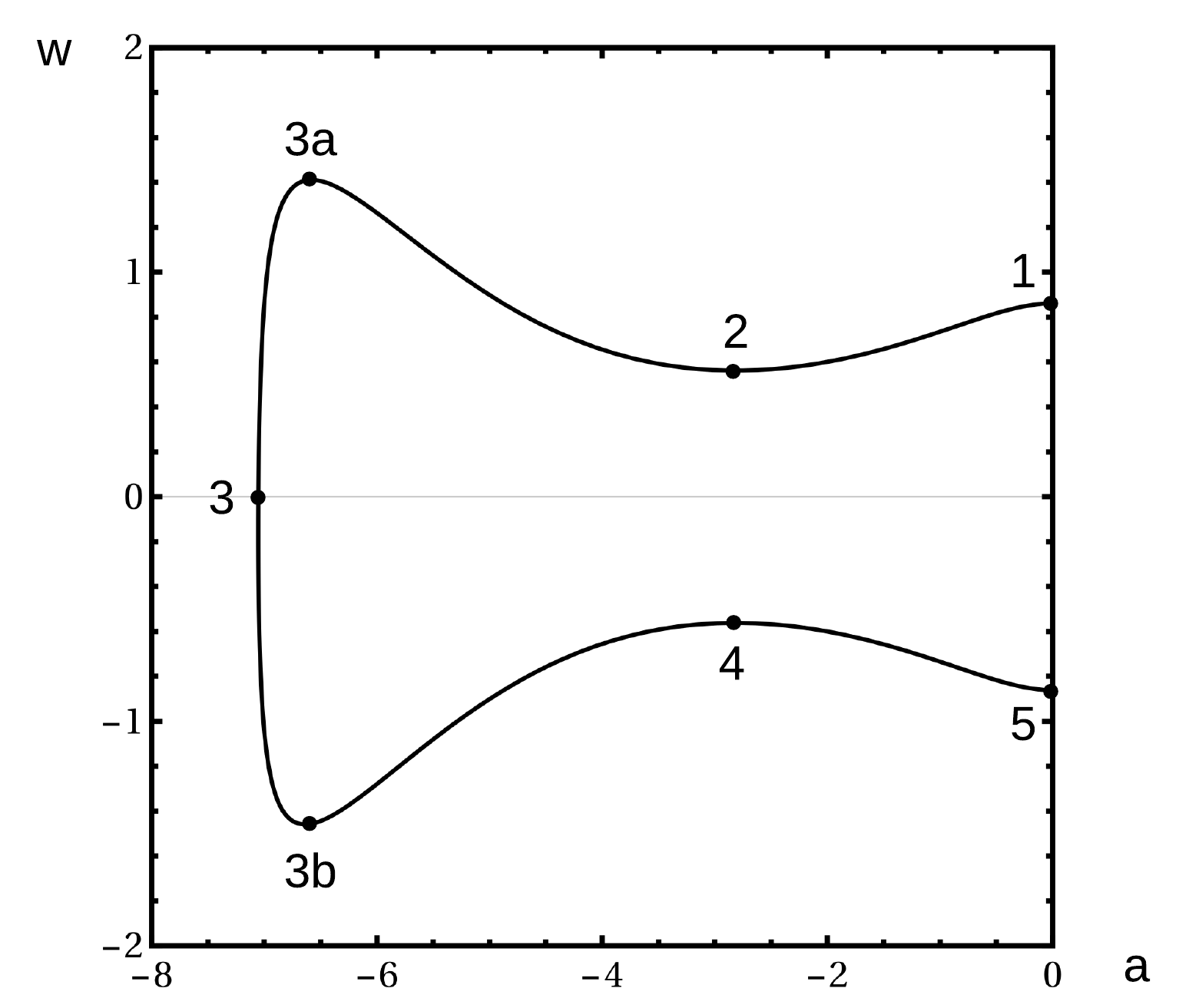}
\caption{On the left: typical RDM solution in logarithmic coordinates, straight lines show contribution of exotic matter. On the right: typical wormhole solution in $a,w$-coordinates.}
\label{f2}
\end{figure}

\paragraph*{Flare-out condition} 
detects the appearance of the wormhole throat, i.e., the minimum of $r$ in the solution. In our model, we identify it as a place where exotic term prevails over RDM term in equation (\ref{daL}). We will consider the region of deeply negative $a$, where this condition can be written simply as
\begin{eqnarray}
&&k_6 - a (k_1-1)/2 + 2k_2 x \gtrsim \log(c_4/2).
\end{eqnarray}

Fig.\ref{f2} left shows typical RDM solution without exotic term. Its main elements have been explained in details in \cite{static-rdm}. Starting from a distant point 1, solution goes through a maximum of $b$ near gravitational radius in point 2 (Schwarzschild region), then rapidly falls down till the minimum of $a$ in point 3 (red supershift region), then goes towards naked singularity with increasing $a$ and decreasing $b$ (ultraviolet region). Physically dimensioned solution are cut off at Planck length before reaching ultraviolet region.

Straight lines on this plot are the lines of constant contribution of exotic term. Variation of $k_6$ shifts this line vertically, while $k_{1,2}$ adjust its slope. If the line is located much deeper than RDM solution (exo1), the contribution of exotic term becomes negligible and RDM solution will be just slightly deformed by it. This situation takes place for deeply negative $k_6$, i.e., small negative $k_3$.

If the line intersects RDM solution (exo2), the exotic term reveals itself. Whether it will lead to the opening of wormhole or not, it can be tested by numerical integration. Our experiments show that the wormhole is opened immediately when the straight line touches RDM solution.  

Considering the equation (\ref{daL}) near the throat $w\to0$, one will have finite values of $a'_L$ only if the numerator in this equation vanishes. For deeply negative $a$ this means $c_4=exo$, coincident with flare-out condition above. Further, considering $w'_L$ equation, one will have a minimum of $r(L)$ only at $w'_L>0$, i.e., $c_4<k_1 exo$. Since $c_4=exo>0$, we have from here an additional restriction: $k_1>1$.

\paragraph*{Structure of the wormhole solution.} Fig.\ref{f2} right shows the obtained solution in $(a,w)$ coordinates. Schwarzschild's maximum of $b$ in the point 2 here corresponds to the minimum of $w$. Fall of $a,b$ in supershift region corresponds to the increase of $w$ till the point 3a, maximum of $w$, minimum of $b$. Then $w$ rapidly falls down, corresponding to $b\to\infty$. The wormhole throat is located in the point 3, $w=0$. After the sign change, on the other side of the wormhole, the solution goes symmetrically through the negative minimum of $w$ in point 3b, Schwarzschild's negative maximum of $w$ in point 4, to the endpoint 5. The plot is constructed for the following values of constants: $\epsilon=0.04$, $c_5=0$, $k_1=2$, $k_2=-1$, $k_6=-11.35$, $r_s=1$, $r_1=100$.

\begin{figure}
\centering
\includegraphics[width=0.49\textwidth]{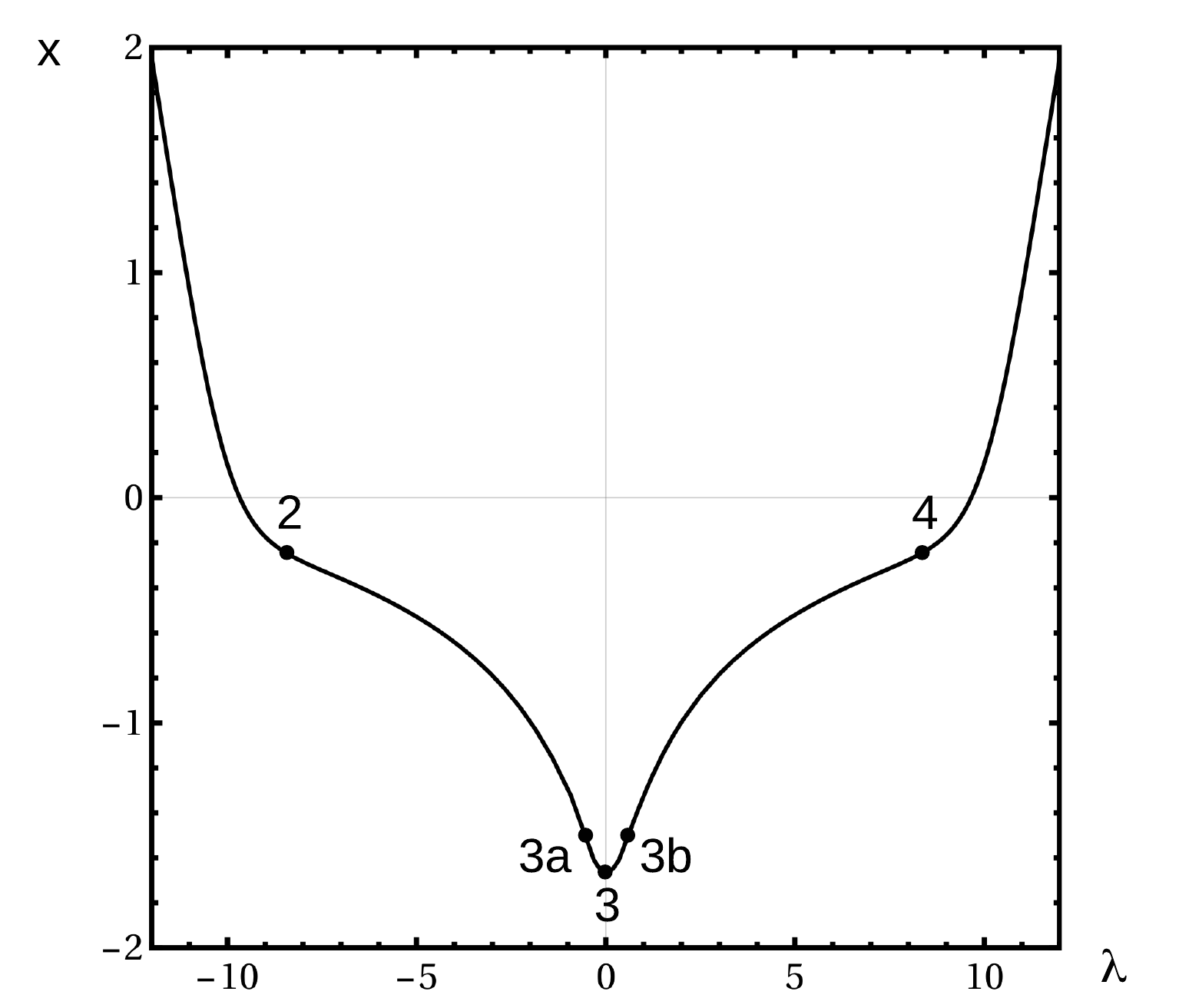}
\includegraphics[width=0.49\textwidth]{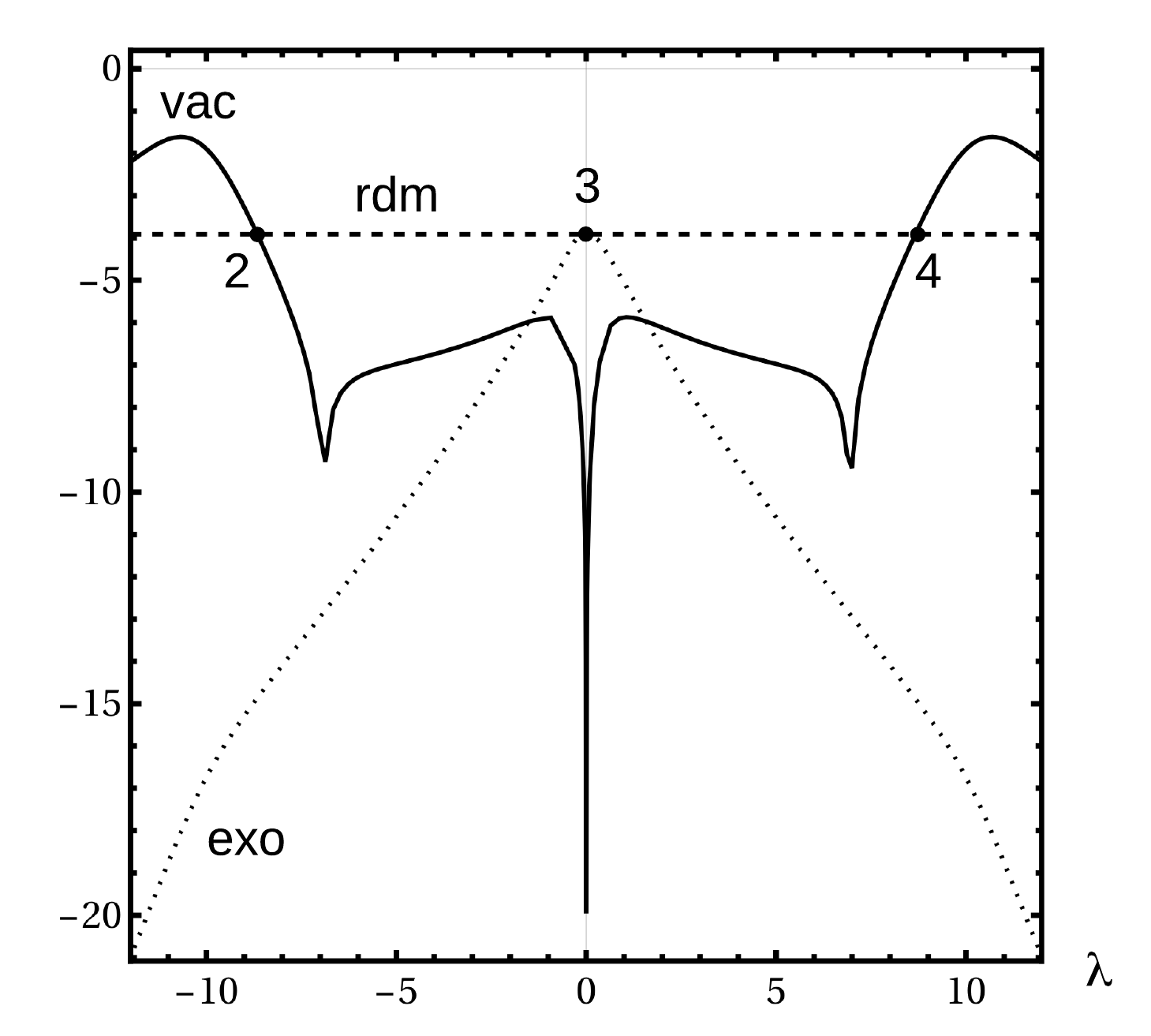}
\caption{On the left: dependence of radius on proper length, in adopted coordinates. On the right: contribution of various terms in the equations.}
\label{f3}
\end{figure}

Fig.\ref{f3} left shows the dependence $r(\lambda)$, where we introduced a transformed proper length variable  $\lambda =-\arcsinh(Lc_\lambda)$. The constant $c_\lambda$ is conveniently chosen to have all features of the dependence visible on one plot. Here $c_\lambda=10^4$. The additive integration constant in $L$ is chosen to have $L=0$ in point 3. The sign of $\lambda$ is selected to have the transition through the wormhole on this plot running from left to right. Since $r'_L=W$, minima and maxima of $w$ correspond to inflection points of $r(L)$, which occasionally coincide with the inflection points of $x(\lambda)$ on this plot. At first one sees a straight segment, corresponding to weak fields and Euclidean dependence $L\sim r$. After inflection point 2, $L$ starts to change much faster than $r$, corresponding to Schwarzschild's large $B$ in $dL=\sqrt{B}dr$. Then, in supershift region,  $r$ changes much faster than $L$, corresponding to small $B$ in this formula. These changes reveal strongly non-Euclidean geometry of solution in the considered region. Further, after inflection point 3a, $r$ goes through the minimum in the throat point 3, then the solution is evolved symmetrically. For constructing this plot we have used smaller $\epsilon=0.02$ and accordingly adjusted $k_6=-17.9$.
This is done to make the variation of this function sharper and its behavior better presented. All dependencies become sharper and sharper when $\epsilon$ is reduced towards physically relevant small values.

Fig.\ref{f3} right shows the behavior of various terms in the equation (\ref{daL}). The first term is the only term remaining after switching off both RDM and exotic contributions, it corresponds to the curvature of empty space-time and is denoted on the plot as 'vac'. The second term is RDM contribution, the third term is exotic one. After initial domination of vacuum term in Schwarzschild's region, RDM term prevails in supershift region and collides with exotic term in the throat of the wormhole. 

In this paper we consider a principal possibility of opening the wormhole in RDM model and feel us are free in the choice of parameters. In particular, we select so negative $k_6$, that the contribution of exotic matter will be hidden deep inside RDM hole and there will be pure RDM solution outside. This allows us to consider the limit of deeply negative $a$ values in the vicinity of the throat and also simplifies our integration procedure. Since the gravitational field of the galaxy in the outside region is well described by RDM term, at least qualitatively, there is no need of cardinal changes there.

\paragraph*{Symmetric wormholes.}
Solutions do not automatically possess a mirror symmetry, visible on the figures. Generally, the solutions on the other side have a different gravitational radius. The gravitational radii can be represented by $r$ values in Schwarzschild's points 2,4, so we have generally $r_2\neq r_4$. In our numerical experiments, for a fixed gravitational radius on entrance, we obtained the gravitational radius on exit, ranging from Planck's length to the size of the universe, dependently on the choice of parameters. 

Also, the redshift function $a$ at the endpoints can differ: $a_1\neq a_5$. Fig.\ref{f1} has shown a special case of redshift function, symmetrically distributed on the wormhole. Generally it is not symmetric and, in particular, it does not necessarily has a minimum in the throat. The displacement of redshift function after the exit from the wormhole can propagate further till the asymptotic region.

As visible in (\ref{asympt}), redshift functions in RDM solutions do not have flat asymptotics, but possess a logarithmic term at infinity: $a=Const-r_s/r+2\epsilon\log r$. Namely the logarithmic term (its derivative) is responsible for flat asymptotics of galactic rotation curves. For physically relevant setting $\epsilon\sim4\cdot10^{-7}$ the logarithmic term at large distances is increasing so slow, that a deviation of the redshift function from the constant remains negligible even at the distances comparable with the size of visible universe. 

On the other hand, dependently on the choice of parameters, the studied wormhole solutions exhibit all possible redshift functions at the endpoint of integration, from infrared to ultraviolet. It corresponds to the asymptotic expansions of the redshift function with different additive constants, i.e., to different rates of time at the opposite sides of wormhole. As extensively discussed in \cite{visser}, this can happen with the wormhole connecting us to the other universe, possessing a globally different time rate (``students of elven folklore will be greatly amused by this observation'', \cite{visser}), 
while in frames of one universe this construction can be used for creating a time machine. 
In less exotic way, the differences of the redshift function at the entries of the wormhole to a certain extent can be explained by different concentrations of mass surrounding the entries. 

Symmetric solution plays here a selected role, since it has automatically equal gravitational radii and equal redshift functions at the entries. Therefore we have tried to achieve this symmetry by adjusting the model constants. While the constant $k_6$ is suitable for adjusting flare-out of the wormhole, the constant $k_2$ can be used for adjusting the symmetry, up to the required precision. For more detailed study of symmetric solutions other methods can be used, explicitly defining $r,a$ as even functions and $w$ as odd function of $L$. This can be achieved by imposing $a'_L=0$ in the throat as additional boundary condition.

\paragraph*{Wormholes: engineering vs modeling.} There is a method for theoretical construction of wormholes, time machines and other extraordinary solutions (``a more engineering oriented approach'', \cite{visser}). 
One sets the desired geometry of space-time and uses field equations to find the necessary distribution of matter. It is similar to the construction of particle accelerator (by a theoretician), taking the desired configuration of electromagnetic field and using Maxwell's equations to derive the necessary charge densities and electric currents. In our model, one can take a ready RDM solution, e.g., the one depicted on Fig.\ref{f2} right, segment 1-2-3a. Take its mirror copy 5-4-3b and close the gap by an arbitrary arc 3a-3-3b. Then use two field equations on $W,A$ to find $\rho_{exo}$, $p_r$ and hydrostatic equation to find $p_t$. In this way one can achieve a compact and even arbitrary thin layer of exotic matter (thin shell, \cite{visser}). There will be no linear EOS and much probably no any EOS whatsoever. The wormhole construction will look more like a particle accelerator above, an artificial facility, where one should maintain definite matter profiles to keep it operating.

In the alternative approach the matter models are fixed from the beginning and the obtained differential equations are solved. Practically, we use an intermediate method, fixing linear EOS for exotic matter and adjusting the constants to obtain the required properties of solutions, just like in engineering approach. 

The model we are using can be further extended, e.g., by considering a polytropic EOS of the form $\rho_{exo}=k_1p_r^{k_7}$. Other models are possible, where the exotic component is not introduced as a separate substance, but as a phenomenon related with the dark matter, e.g., self-interaction. This phenomenon should appear only at high densities of dark matter and should give a contribution to energy-momentum tensor with the required sign. It should be equivalent to negative additions to density and pressure, as necessary for opening the wormhole.

\section{Numerical integration}\label{sec3}

We use {\it Mathematica}, algorithm {\tt NDSolve} for numerical integration of differential equations. Transition to the variables $a,x,w$ is similar to the logarithmic transformations, used in \cite{static-rdm}. They reduce stiffness of the system and make numerical procedure more stable. 

The difference is that $w$ is now defined by $\arcsinh$ transformation. It allows for transition to the other side of the wormhole through zero value of $W$. Simultaneously, it protects the integrator from singularities at large $W$, i.e., extremely small $B$ in supershift region. Earlier the variable $b$ defined by $\log$ transformation was used, which did not allow for zero values of $W$, while it was similarly protected against small $B$ values. 

Asymptotic region of large $r$ earlier corresponded to $B\sim1$, i.e., $b\sim0$. Now it corresponds to a finite non-zero value $w=\arcsinh 1$. As a result, new integrator is less suitable for integration of small terms across large distances, happening in the asymptotic region. This integration will add small terms over essentially non-zero value and the error will be rapidly accumulated. Practically, it is better to start integration on the border of Schwarzschild region 1b and stop it in the mirror point 5b after exit from the wormhole. As a starting point, one can take ISCO radius $r_{1b}=3r_s$. In this point non-trivial curvature of space-time is already achieved and the integrands become sufficiently large. Integration outside of this region can be performed with the old integrator \cite{static-rdm}. One can omit exotic terms in the equations or not, it is only important that $b$ should be among the integration variables. At our setting of parameters, the contribution of exotic term is negligible everywhere at $r>r_{1b}$. 

The system (\ref{daL}) defines a vector field in 3D space: $(x,a,w)'_L=(v_x,v_a,v_w)$. It is similar to 2D vector field $(a,b)'_x=(v_a,v_b)$ defined in \cite{static-rdm}, but now the integration parameter is $L$ and the field explicitly depends on $x$. The system is still autonomous, does not depend on integration parameter $L$.

The integration of this field can be done as follows:
\begin{eqnarray}
&&x'_y=v_x/norm,\ a'_y=v_a/norm,\ w'_y=v_w/norm,\\
&&norm=\sqrt{v_x^2+v_a^2+v_w^2},\nn
\end{eqnarray}
i.e., by transforming to a new integration parameter and applying a normalization to the field. This method is a generalization of the switching scheme used in \cite{static-rdm}, proceeding from $da/db=f(a,b)$ to $db/da=f^{-1}(a,b)$ whenever $f$ becomes too large. Now large terms in the equation automatically adjust new integration parameter $y$ to the variation of fastest variable in the system. This transformation helps standard step adaptation algorithms, used in {\tt NDSolve}, and leads to more stable integration. 

The proper length variable can be easily found by post-integration:
\begin{eqnarray}
&&L=\int dy/norm,
\end{eqnarray}
the integration constant here can be selected to provide $L=0$ in the throat of the wormhole.

Complexity of wormhole scenarios leads to increased integration time, from 0.006~sec for pure RDM solutions to 3~min for wormhole solutions, on 3GHz CPU. Precision of integration can be estimated similar to \cite{static-rdm}, artificially reducing the step size by a factor of 10 and comparing the results. Precision remains similar, for the Milky Way scenario considered below, integrated in the region 1b-5b, the values of $a,w$ in wormhole zone have an absolute error $\sim10^{-1}$, while in relation to the amplitude of variation $\sim10^6$ of these variables in this zone it comprises the relative error $\sim10^{-7}$.

\section{Matching the model with a real galaxy}\label{sec4}

The results of integration with parameters of Milky Way galaxy are shown in Table~\ref{tab_mw}.

The integration starts from pure RDM solution at the beginning of Schwarz\-schi\-ld zone at $r_{1b}=3.33\cdot10^{10}$m. Starting values $a_{1b}$,$b_{1b}$ are taken from \cite{static-rdm}, $b_{1b}$ is converted to $w_{1b}$ value.

Further, at $r_{2}=1.11\cdot10^{10}$m, $w$ reaches a minimum, corresponding to a maximum of $b$ at the end of Schwarzschild zone. After that, typical for RDM solution supershift zone starts, with decreasing $a$ and increasing $w$.

The exotic matter reveals itself near $r_{3}=1.28\cdot10^8$m. In the point 3a, located a bit higher than $r_{3}$, the value $w$ passes a maximum and rapidly falls down. It passes through zero in the point 3, the wormhole throat, then goes to the negative side. After that the solution evolves in approximately symmetric way.

The symmetry is not exact, in particular, the gravitational radius $r_{4}$ on exit is 30\% larger than $r_{2}$ on entry. There is also a small deviation of the redshift function between the start and the end of integration: $a_{5b}-a_{1b}=-5.76\cdot10^{-5}$. Integration outside of this region shows that the shift between the curves becomes constant, while the numerical value remains on the same level: $\Delta a=-5.78\cdot10^{-5}$.

After variation of a single parameter $k_2=-56.11$ we obtain a solution with the gravitational radius on exit 50\% less than on the entry and the redshift of opposite sign: $\Delta a=+5.57\cdot10^{-5}$. Thus, the symmetric solution is bracketed and one can find it with a better precision, if required.

\begin{table}
\begin{center}
\caption{Wormhole solution in RDM model of Milky Way galaxy}

\def\arraystretch{1.1}
\begin{tabular}{|c|c|}
\hline
&$\epsilon=4\cdot10^{-7}$, $r_s=1.2\cdot10^{10}$m,\\
model parameters&$c_4=3.77124\cdot10^{-7}$, $c_5=1,$\\
&$k_1=2$, $k_2=-51.453$, $k_6=-6.609\cdot10^5$\\ \hline
begin of Schwarzschild zone&$r_{1b}=3.33\cdot10^{10}$m, $a_{1b}=-b_{1b}-2.06\cdot10^{-5}$\\ 
(start of integration)&$b_{1b}=0.404$, $w_{1b}=\arcsinh e^{-b_{1b}/2}$\\ \hline
end of Schwarzschild zone,&$r_{2}=1.11\cdot10^{10}$m, \\ 
begin of supershift&$a_{2}=-14.81$, $w_{2}=1.228\cdot10^{-3}$\\ \hline
end of supershift,&$r_{3a}=r_{3}+2.1\cdot10^{5}$m, \\
wormhole mouth&$a_{3a}=a_{3}+1.05$, $w_{3a}=6.628\cdot10^{5}$\\ \hline
wormhole throat&$r_{3}=1.28\cdot10^8$m, $a_{3}=-1.326\cdot10^6$, $w_{3}=0$\\ \hline
mirror wormhole mouth,&$r_{3b}=r_{3}+4.2\cdot10^{5}$m, \\
end of mirror supershift&$a_{3b}=a_{3}+0.77$, $w_{3b}=-6.628\cdot10^{5}$\\ \hline
begin of mirror supershift,&$r_{4}=1.42\cdot10^{10}$m, \\ 
end of mirror Schwarzschild zone&$a_{4}=-14.87$, $w_{4}=-1.229\cdot10^{-3}$\\ \hline
begin of mirror Schwarzschild zone&$r_{5b}=4.27\cdot10^{10}$m, $a_{5b}=-b_{5b}-7.82\cdot10^{-5}$,\\ 
(end of integration)&$b_{5b}=0.404$, $w_{5b}=-\arcsinh e^{-b_{5b}/2}$\\ \hline
\end{tabular}
\label{tab_mw}
\end{center}
\end{table}

\section{Conclusion}\label{concl}
We have studied a possibility that a massive dark object in the center of galaxy is a wormhole, described by RDM model. Wormhole solutions require a contribution of exotic matter, for which we have considered several options. The exotic matter in a form of a perfect fluid with a linear anisotropic equation of state creates wormhole solutions with the necessary asymptotics. The flows of dark matter pass through the wormhole and form a mirror galaxy on the other side. We have studied the influence of model parameters to the shape of solution. One of the parameters, $k_6$, influencing the total amount of exotic matter, enters additively to flare-out condition and regulates creation of the wormhole. The other parameter, $k_2$, entering the equation of state, can be used to regulate the symmetry of the wormhole. The solution matching parameters of Milky Way galaxy can be dimensioned to have the wormhole located deep inside the central massive dark object, by a factor of hundred below its gravitational radius.

\end{document}